\documentclass[aps,prl,
showpacs,showkeys,amsmath,amssymb,
groupedaddress]{revtex4}
\usepackage{graphicx}
\usepackage{epsf}
\input epsf
\def\ll{\label}
\def\re{\ref}

\def\r1{(\ref{$1})}

\def\ba{\begin{array}{c}}

\def\ea{\end{array}}

\def\l{\left}
\def\l({\left(}
\def\r){\right)}
\def\r{\right}

\def\la{\lambda}

 \def\be{\begin{equation}}
\def\bc{\begin{center}}
\def\ec{\end{center}}
\def\bit{\begin{itemize}}
\def\eit{\end{itemize}}
\def\ee{\end{equation}}
\def\ed{\end{document}}
\def\bea{\begin{eqnarray}}
\def\eea{\end{eqnarray}}
\def\efr{\end{flushright}}

\begin{document}

\title{
 Modelling rogue waves through exact dynamical
 lump soliton controlled by  ocean currents}
\author{ Anjan Kundu} 
\email{anjan.kundu@saha.ac.in}\author{Abhik Mukherjee}
\email{abhik.mukherjee@saha.ac.in}  
\author{Tapan Naskar}
\email{tnaskar@gmail.com} 
\affiliation{Theory Division,
Saha Institute of Nuclear Physics\\
 Kolkata, INDIA}
\begin{abstract}
 Rogue waves are extraordinarily high and steep isolated waves, which appear
suddenly in a calm sea and disappear equally fast.  However, though the
Rogue waves are localized surface waves, their theoretical models and
experimental observations are
 available  mostly in one dimension(1D) with the majority of them admitting
only limited and fixed amplitude and modular inclination of the wave.  We
propose a two-dimensional(2D), exactly solvable Nonlinear Schr\"odinger
 equation(NLS) , derivable from the basic hydrodynamic equations and endowed
 with integrable structures.  The proposed 2D equation exhibits modulation
 instability and frequency correction induced by the nonlinear
effect, with a directional preference, all of which can be determined through
precise analytic result.  The 2D NLS equation allows also an exact lump
solution which can model a full grown surface Rogue wave with adjustable
height and modular inclination.  The lump soliton under the influence of an
ocean current appear and disappear preceded by a hole state, with its
dynamics controlled by the current term.
  These desirable
properties make our exact model promising for describing ocean rogue waves.
\end{abstract}
 \maketitle

\section{1. Introduction}

 The mysterious  ocean rogue waves (RWs) are  
 reported to being observed in a relatively calm sea, where they, as a
localized and isolated surface waves, apparently appear from nowhere, make a
sudden hole in the
 sea just before attaining surprisingly 
 high amplitude   and  disappear    
 again without a trace
{\cite{BBCnews10,rogRev,prldeter11,prlCurr11,Zakh09}}.
 This elusive freak wave  caught the 
 imagination of the broad scientific
community quite recently
 {\cite {Nat07,prlacust08,prl2DWexp09,prlCav09,
Natphys10,prlcapil10,prlwTank11,prlmulti11}}, 
 triggering off an upsurge in
theoretical 
 {\cite{
 prl1dPnls00,pla2Dnls00,prl2Dnls10,Zakh09}} 
 and experimental 
 \cite{prldeter11,Nat07,prlacust08,prl2DWexp09,
prlCav09,Natphys10,prlcapil10,prlwTank11,prlmulti11} 
 studies of
 this unique  phenomenon.
 For identifying such extreme waves
 the suggested
signature of these rare events 
  is a  deviation of 
 the probability distribution function (PDF) of the
 wave
 amplitude from its usual random Gaussian
 distribution (GD),  by having   a  long-tail,  indicating that the
appearance
 of high
 intensity pulses more often, has much higher probability 
 than that
predicted by the GD
 {\cite{AkhmDeb09}}.
 In conformity with this definition
RWs were detected 
 in a photonic crystal fiber {\cite{Natphys10}}, in a multi-stable state
 of an
 erbium doped fiber laser
 {\cite{prlmulti11}}, in chaotic but
deterministic regime
 of optical injected semiconductor
 lasers 
{\cite {Nat07,prldeter11}}, 
  in nonlinear optical cavity {\cite{prlCav09}}, 
in acoustic turbulence in He II {\cite{prlacust08}}
 and other set ups  {\cite{prlcapil10}}.  On the formation of the ocean RWs
a number of supporting theories have been developed {\cite{rogRev}}.  Among
various possible factors contributing to the creation of the RW, the
modulation instability (MI) supported by the nonlinear effect is believed to play
a crucial role, by inducing preliminary amplification of water wave height ,
which may trigger self attractive nonlinear interaction, initiating the RW
 formation {\cite{Ruban}}.  The 
MI can also cause
wave-wave interaction leading to the four-wave mixing at matching
frequencies and wave numbers, inducing resonance effect which might also
develop into a RW {\cite{Zakh09,prlcapil10,ZakhJETP05}}.  Since the
four-wave nonlinear interaction, a leading order
 nonlinear effect in deep-sea waves, is found also to be a dominant
interaction in the  nonlinear Schr\"odinger (NLS) equation
 \be iq_t=
q_{xx}+2|q|^2q, \ll{nls} \ee 
with the subscripts denoting partial differentiation, 
the NLS based nonlinear models 
 are the most accepted ones for the RW, though often with certain
modifications to include higher oder dispersion or ocean currents, which 
are suspected to have a deciding role  in the formation of the RW 
\cite{prlCurr11}.
 In extended space dimensional systems the nonlinear effect due to the 
MI in combination with a space-asymmetry, directional
spectra and broken symmetry due to nonlocal coupling is suspected to be the
major causes of such extreme waves  
 {\cite{prl2DWexp09,prlCav09,prl2Dnls10}}.
The NLS equation (\re{nls}) is a well known evolution equation with integrability properties like having a Lax pair
 and exact soliton solutions {\cite{solit}}. Some
models of RW generalize the NLS equation
 with the addition of extra terms 
on physical grounds, like ocean current 
 {\cite{prlCurr11}}, nonlinear
dispersion 
 {\cite{prl1dPnls00,plaHomo08}} etc. However such modifications of the NLS
equation (\re{nls}) make the system nonintegrable, allowing only
 numerical  solutions.  The most popular 1D RW model  is a unique
analytic rational solution of the original  NLS equation (\re{nls})
\cite{peregrin83}, given by the
Peregrine  breather (PB) \cite{Natphys10,pla2Dnls00,prlwTank11}
 or its higher order versions \cite{MrogPRE09}-\cite{RWtriple11} and the
trigonometric variants \cite{ma,akhmed}.  However, since the RW is an
aperiodic event with a single appearance, the trigonometric breather
solutions, due to their periodic nature, are not much suitable for a direct
description of the RW.  Nevertheless, interestingly these breather solutions,
periodic in time \cite{ma} or in space \cite{akhmed}, degenerate to the
rational PB solution (\re{Pbr}) at their periods going to
infinity\cite{Dysthe}.

Note that the conventional
soliton solution 
 of the NLS equation (\re{nls}), representing 
 a
 localized translational wave
behaves like a stable particle and unlike a RW   propagates 
   with   unchanged shape and  amplitude. Tsunami waves, though highly
devastating, also exhibits different nature than the ocean RW.  The ocean RW
are deep sea waves with 2D character, localized in both space dimensions and
appears as a single-peak event for a short interval of time.  Tsunami waves
on the other hand are manifested only in shallow water near the sea shore,
though generated in the deep sea and propagate across a long distance. 
In the deep sea tsunami waves behave like 1D translatory wave, moving very
fast with insignificant amplitude{\cite{TsuLax}}.
Therefore tsunamis and the RWs exhibit different features and
dynamics and need different types of modelling which for the RW is still an
open problem.
  More details on the
progress in the study of the ocean RW can be found in some excellent reviews
on the subject \cite{rogRev}.

\subsection{1.1.  RW model on a 1D line}
   In contrast to the soliton or the trigonometric breather solutions of the 
NLS equation  (\re{nls}), its  exact rational PB solution 
\bea q_P(x,t)&= &e^{-2it}(u+iv), \ u=G-1, \ v=-4tG, 
\nonumber \\ \mbox{ 
 where
}
\    G&=& 1 /F(x,t), \ F(x,t)=x^2+4t^2+\frac 1 4, 
\ll{Pbr} \eea
 represents a 
breather mode  
  with unit amplitude at both   
   distant past and  future. The  
  amplitude of the wave rises suddenly  at   $t=0 $,
 attaining  its maximum at $x=0  $, though      
subsiding  with time
again  to the same  breathing  state.
This   intriguing  behavior makes  the PB  a popular
candidate for the RW
{\cite{Natphys10,pla2Dnls00,prlwTank11}}. 

Since the characteristics of the envelop wave is the most significant in the
description for the RW, the modulus of the PB solution (\re{Pbr})
\begin{equation} |q_{P}(x,t)|= (u^2 + v^2)^\frac{1}{2} =[(G-1)^2 + (4 t
G)^2]^\frac{1}{2},\label{Pbmod} \end{equation} with $G$ as in (\re{Pbr}), is
used in describing the RW profile.  The full grown 1D RW at $t=0 $ therefore
may be represented by \begin{equation} |q_{P}(x,0)|=
(G-1)|_{t=0}=[\frac{1}{x^2 + \frac{1}{4}}-1],\label{Pbr0} \end{equation} as
shown in Fig.  1.  The maximum amplitude as seen from (\re{Pbr0}) is
attained at $x=0$ as $|q_{P}(0,0)|=3$.
 The modular inclination defined as 
\begin{equation}
 S_{P}^x(x)= \frac{\partial}{\partial x}
 |q_{P}(x,0)|=-\frac{2 x}{(x^2 + \frac{1}{4})^2}
\label{PBincl}\end{equation}
attains
 its maximum $S_{Pmax}^x(x_{m})= 3 \sqrt{3}$ at $x_m= \pm
\frac{1}{2\sqrt{3}}.$


\begin{figure}[!h]
\centering
{
 \includegraphics[width=4cm, angle=-90]{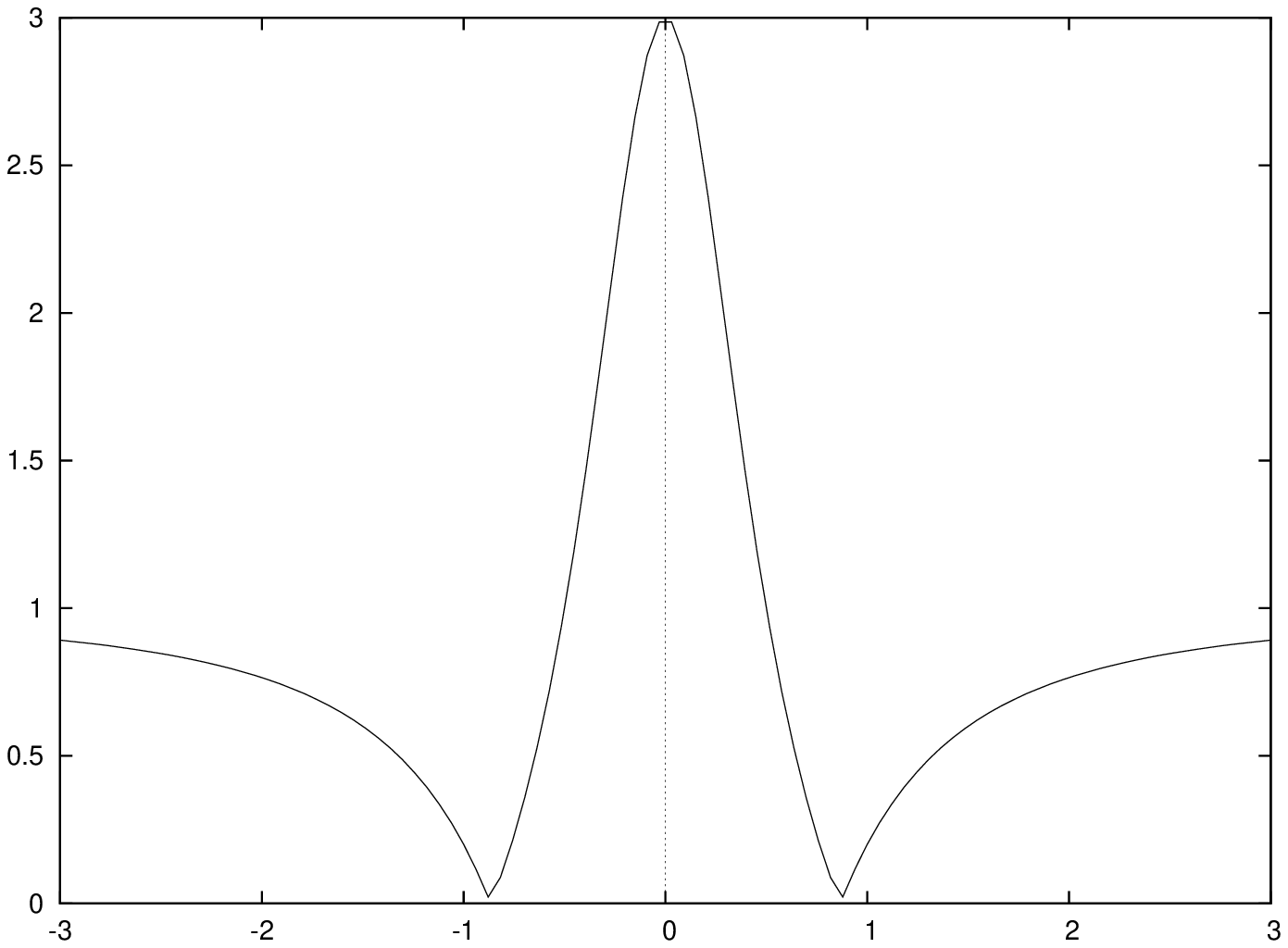}
}

\caption{ Amplitude variation of the full grown 1D rogue wave,
  modelled by the  modulus of the 
static Peregrine breather $|q_P(x,0)| $ .
 The maximum amplitude $3$ is
  attained at $x=0$,  while  it 
goes to  its asymptotic  value $1$  at $x \to \pm \infty $. The  maximum 
inclination attainable is  $3 \sqrt 3$ at $x=\frac{\sqrt{3}}{6}$,
 and becomes $0$ both at
$x=0$ and   $x \to \pm \infty $.
}
\label{fig:1}
\end{figure}


Notice  however that, 
the NLS equation (\re {nls}) together with its different generalizations   
are   equations  in  $(1+1) $-dimensions
and therefore  all of  their   solutions,  including the  PB and its 
higher order generalizations,  
 can describe  the
 time evolution of a wave only along an one dimensional
 line
 (as in Fig.  1).
 Looking more closely into the PB
 we also realize that
 the maximum amplitude of this solution describing an
1D RW
  is fixed, and  just 
three times that of  the background waves (see Fig. 1).
The modular inclination of this wave 
as well as the fastness
  of its  appearance 
 are also fixed,
since solution (\re{Pbr}) admits
 no free parameters. This situation can be improved to obtain higher
amplitude and modular inclination of the PB model by using higher order
rational solutions \cite{MrogPRE09}.  For example, the next higher order PB
known also as Akhmediev-Peregrine breather can enhance the maximum wave
elevation
 by a factor of five, while the next one by a factor of seven and
so on, with an intriguing enhancement of factors by increasing odd numbers.
 Such increments in amplitude however are discrete and could be
achieved at the cost of going to new solutions
 with increasingly complicated
structures involving higher and higher order polynomials
 \cite{MrogPRE09}. 
  The maximum  amplitude  and modular inclination reachable by this class of
solutions are  fixed 
 due to the absence of relevant  free tunable
parameters, making it 
 difficult to adjust to 
 the continuously varied range of shape and sizes of the
  observed  oceanic RWs. However recently higher order rational solutions to
the NLS equation allowing free parameters have been
discovered {\cite{Dubard,RWtriple11}}, though they seem to represent 
multi-peak wave in the $x-t$ plane for the nontrivial choice of
parameters {\cite{RWtriple11}. The single-peak solution which is suitable for
describing RW having a single appearance in time, is obtained unfortunately
for a trivial choice of the free parameters.  The trigonometric breathers
\cite{ma,akhmed} also contain free
 parameters  \cite{Dysthe}, though such
periodic solutions, as mentioned already, are different in nature than the
single crest RW event.
 The  crucial fact however is that, the 1D spatial nature remains the
  same for the whole class of the PB
solutions, including its higher order rational and 
trigonometric generalizations. 
Therefore modelling an ocean RW, which is a 
 2D surface wave, by this class of 1D PB solutions  remains problematic. 
\subsection{1.2. Need for a  RW model on a 2D plane}

Therefore, though the well accepted class of   PB or
   other  solutions   of the 
 generalized NLS equation could fit into the working definition of the
ocean RW, saying any wave with height more than twice the nearby
 {\it significant height}
 (average height among one-third of the highest waves) could be treated as
the RW {\cite{AkhmDeb09}}, they perhaps, with their  restricted
characteristics, can  explain successfully only fixed and 
moderately intense RW-like events on a 1D line,
as observed
  in  water channels {\cite{prlwTank11}}, optical fibers
{\cite{Natphys10,prlmulti11}}
 or   optical lasers {\cite{prldeter11,Nat07}}, but seem to be
  not satisfactory  for   modelling  the ocean surface  RWs.
 
Oceanic RWs are said to be consist of an almost vertical wall of water
preceded by a trough so deep, that it was referred to as a hole in the sea
{\cite{Rogwiki}}.  In march 2001,  two reputed ships named as Bremen and
Calendonian Star, carrying hundreds of tourists across the South Atlantic,
had a devastating encounter with RW like events.  It is reported by the
witnesses that a giant isolated wave of around 30 meter high, fell upon the
ship like a wall of water, out of no where and disappeared again without a
trace \cite{BBC2}.
At the initiatives of 11 organizations involving several countries in EU
tasked the Earth - scanning satellites, named ERS-1 and ERS-2, to send images
from a localized area of $10\times 5 $ $km^2$ on the sea surface at certain
locations to spot the
 possible occurrence of rogue waves \cite{BBC2}.

All these available facts and information suggest that unlike the tsunami
and internal waves, pictures of which can be seen through satellite images
\cite{nasa}, ocean RWs with the hole states must have a 2D character,
localized in both the space dimensions.
 In 2D water basin
 experiments as well as in the related
simulations the amplitude and
 the modular inclination of
 the RWs were found to be  higher
{\cite{prl2DWexp09,prlcapil10,prl2Dnls10,pla2Dnls00}}
 than those predicted and observed in 1D {\cite{prlwTank11,pla2Dnls00}}.

The above arguments should be convincing enough to go beyond the 1D
equations and search for a suitable (2+1) dimensional equations, to find 
 a 2D alternative to the PB and other solutions of the 1D NLS 
 equation,
for constructing a more realistic model for the ocean RWs.
 
There are many nonlinear equations known in (2+1) dimensions having fruitful
applications in various fields.  Some of them allow exact analytic
solutions, while others permit only approximate numerical simulation.

The well known KP equation is an integrable extension of the KdV equation
to 2D space {\cite{KP}} describing the dynamics of a real field. 
 However the KP
equation like the KdV is a shallow water model, whereas the RW is naturally
a deep water phenomenon.

There are also several equations extending the 1D NLS equation to (2+1)
dimensions.  From the basic hydrodynamic equations, by taking the
perturbation analysis to a higher order, Dysthe
 has derived for the deep water waves a 2D evolution equation 
\cite{Dysthe}}. The Dysthe equation in general is non integrable .

The Davey-Stewartson equation \cite{DS} is an integrable generalization
where the existence of rogue wave  has been analyzed  \cite{Ohta}. 
However such rogue wave solutions are reducible to the PB solution by a
simple rotation in the plane.  BLP equation \cite{BLP} is another $(2+1)$
dimensional integrable equation, defined through two real coupled equations. 
Recently a RW type solution has been found in this equation allowing a free
parameter \cite{MuDaiZhao}.  However since the BPL equation describes wave propagation along
a channel , its applicability in modelling the ocean RW is questionable.

Zakharov have proposed several 2D equations, some of them are integrable
\cite{ZakhJETP72,Zakhar} while others are not \cite{ZakhJETP05}.  Though
these equations
 are applicable in other fields, the model proposed in \cite{ZakhJETP05}
 seem
to be a successful model for the RW.

A straightforward 2D extension of the NLS equation :
\be  i  q_t= d_{1} q_{xx} -d_{2} q_{yy} +2|q|^2q,
 \ll{2dnls} \ee where $q(x,y,t)$ is a slowly varying envelop and $d_{1}$,
$d_{2}$ represents linear dispersion coefficients (\cite{prl2Dnls10}) of the
deep water gravity wave,  was proposed in connection with RW
{\cite{prl2Dnls10,pla2Dnls00}}.

 Note however that the 2D NLS equation
(\re{2dnls}) is not an
 integrable system and  gives  only approximate numerical solutions with no
stable soliton.
 Nevertheless,  this unlikely candidate  is found to exhibit RW like
structures
 numerically, with higher amplitude and modular inclination and with an
intriguing directional preference {\cite{prl2DWexp09,prl2Dnls10}}
 with  broken  spatial symmetry 
 {\cite{prlCav09,pla2Dnls00}}.  However though the experimental and
theoretical studies on nonlinear systems in 2D space have shown promises in
describing more realistic situations in the formation of 2D ocean RWs,
unfortunately, all of them can give only approximate numerical results and
most of this models could not consider the effect of ocean current which is
supposed to play a crucial role in the formation of ocean RWs 
{\cite{pla2Dnls00,prl2Dnls10}.

\section{2. Proposed  integrable 2D NLS equation}

 In the light of not so satisfactory present state in modelling the deep sea
RWs, we propose an {\it integrable}
 extension of the  2D NLS equation:
 \bea  i  q_t&= &  d_{1}q_{xx} -d_{2} q_{yy} +2iq(\sqrt{d_{1}}
j^x-\sqrt{d_{2}}j^y), 
 \ \ j^a \equiv qq^*_a-q^*q_a,
 \ll{2dInls} \eea
 allowing an exact lump-soliton as a suitable RW model.  In
(\re{2dInls}) the linear dispersion relation is exactly same as the
conventional water wave dispersion as described in (\re{2dnls}), with the
only difference from this well known 2D NLS equation being in the nonlinear
term.  Notice that, when the conventional {\it amplitude}-like nonlinear
term
  in the non-integrable equation (\re{2dnls})
is replaced by a  nonlinear {\it current}-like term (expressed through $j_x,j_y$), the resulting equation 
(\re{2dInls}) miraculously becomes a completely
integrable system with all its characteristic properties,
 which is much 
 rarer in 2D than in 1D.
Before proceeding further  observe, that 
 through  scaling and a $\frac \pi 4$ rotation on the plane 
: $ (x,y) \to (\bar x,\ \bar y) $ with
$ \bar x=\frac 1 2(-\frac{x}{\sqrt{d_{1}}}+\frac 1 {\sqrt{d_{2}}}y),\ 
\bar y=\frac 1 2(\frac{x}{\sqrt{d_{1}}} + \frac 1 {\sqrt{d_{2}}} y)  $ 
and $ \bar t= 2t$,   
our 2D NLS equation (\re{2dInls}) can be simplified  to 
 \be  i  q_t+q_{ x  y}  +2iq(qq^*_{x}-q^*q_{ x})=0, \ll{2dsInls} \ee
 where the $bar $ over the coordinates is omitted.  Encouragingly, our 2D
NLS equation (\re{2dsInls}), at par with the well known 1D NLS equation
 is derivable from the more fundamental hydrodynamic equations and exhibits
 MI together with a  nonlinear frequency correction,
as we show below.
 Equation  (\ref{2dsInls}) admits also  exact 
 soliton and breather solutions through the standard formalism of Hirota's
bilinearization and an associated Lax pair as well as  an infinite set of
conserved charges  \cite{arXiv12}, proving thus the integrability of this 
nonlinear equation .  
 More satisfactorily,  equation 
(\re{2dsInls}), as we see below,  admits an exact 2D generalization
 of the PB with 
 the desirable properties of a realistic surface RW. It is promising that
many characteristic properties like directional preference, MI, appearance
 of higher amplitude etc observed theoretically and
experimentally in connection with the formation of RW in 2D
models \cite{{prl2DWexp09,prlCav09,pla2Dnls00,{prl2Dnls10}}}, 
which remained as numerical approximations, get
confirmed through analytic result in our model based on the integrable
equation (\ref{2dsInls}).
 
\subsection{2.1.  Nonlinear frequency correction and modulation instability}
Instability of a planer wave, appearing due to the interplay between
dispersion and nonlinear effect called Benjamin Feir or MI \cite{BF},
 which has been in the continuous focus for
many years \cite{McLean,KharifRamma},
 has gained more importance recently in the context of the RW. 
The nonlinearity
 and the MI are supposed to be the basic reason behind
the formation of RWs.  Therefore,
 before
progressing further with our 2D NLS equation (\re{2dsInls}),
 we focus on the  correction of its linear  frequency induced by  the
nonlinear effect and the appearance of the MI mediated
by such nonlinearity in the system.  For investigating the contributions to
the frequency due to the linear dispersive and the nonlinear term in
(\re{2dsInls}), we insert the plane wave solution $ q_0= A_0 \
e^{i(\omega t+k^xx+ k^y y)} ,$ with $A_{0}$
 as the real constant  amplitude, $ \omega$ as frequency and
$(k^x,k^y) $ as the wave vector .  For the plane wave to be
 an exact solution of (\ref{2dsInls}), the frequency should be $ \omega= \omega_L+\omega_{NL},
\ \omega_L=-k^xk^y, \ \omega_{NL}=4A_0^2k^x,$
 where $ \omega_L $
is the frequency due to linear dispersion and $\omega_{NL} $ is its
nonlinear correction, which depends on the amplitude  of the wave as well as
on the x component of the  wave vector.

Now to explore  the onset of MI  in the system
affecting this plane wave solution, we perturb it by a small parameter
function  $\epsilon(x,y,t) $. Note that the perturbation is considered in both the space directions
since its importance in the instability in 2D is emphasized in the context
of RW formation \cite{pla2Dnls00}.
 The solution \bea q_\epsilon= ( A_0 +\epsilon)\
e^{i(\omega t+k^xx+ k^y y)} , \ll{Aep}\eea
 neglecting the higher order terms in $\epsilon $  yields from (\re{2dsInls}) a linear
equation for $\epsilon $ as \bea  i\epsilon_t+ \epsilon_{xy}+
i(k^y\epsilon_{x}+k^x\epsilon_{y})+
2 i
A_0^2(\epsilon_{x}^{*}-\epsilon_{x} )+4A_0^2 k^x(\epsilon^*+\epsilon)=0
. \ll{epEq}\eea
The appearance of the last two terms in equation (\re{epEq}) is due to the
nonlinearity.  For detecting the
 instability of the perturbation we represent
 $\epsilon= c_1 e^{i(\omega_{m}t+k^x_{m}x+k^y_{m}y)}+
 c_2 e^{-i(\omega_{m}t+k^x_{m}x+k^y_{m}y)}$
 Inserting this form of perturbation in equation (\re{epEq}) and arranging the independent
terms we get a set of two homogeneous equations for the arbitrary
coefficients $c_1,c_2, $ nontrivial solutions of which can exist only when
the determinant  of the matrix vanishes leading to the necessary
relation
 $ { \bar \omega_m}^2=K^2- {\Omega
_c} , $\ 
{where }\
$
  \bar \omega_m= \omega_m -\omega_{0},$ \ and  
\ $ \omega_{0}=2A_0^2 k_m^x-{k^x}{k_m^y}-{k_m^x}{k^y} ,$
\ $ K= {k_m^x}{k_m^y}-4A_0^2 k^x, \  \ \Omega_c=
4A_0^4(4 {k^x}^2 -{k_m^x}^2),$
which gives finally
\bea
 \omega_m=  \omega_0 \pm i \omega_{I} 
 ,\
 \omega_{I} = (\Omega_{c}-K^2 )^\frac{1}{2}.
\ll{growthrate}
\eea
Therefore,  under the condition $K^2< {\Omega_c} $ with
  ${\Omega_c}>0 $, i.e when $|{k_m^x}|<2| {k^x}|$ the modulation frequency $
\omega_m$ can acquire an imaginary part $\omega_{I}, $ initiating an
exponential growth of perturbation with time $t$ and hence onsetting the
MI.  $\omega_{I}$ is the growth rate of the instability
given by (\ref{growthrate}),
a graphical form of which is presented in Fig.  2,  showing its dependence on
the longitudinal and transverse directions through $k_{x}^m$, and $k_{y}^m$
respectively.

\begin{figure}[!h]
\centering
{
 \includegraphics[width=5cm, angle=0]{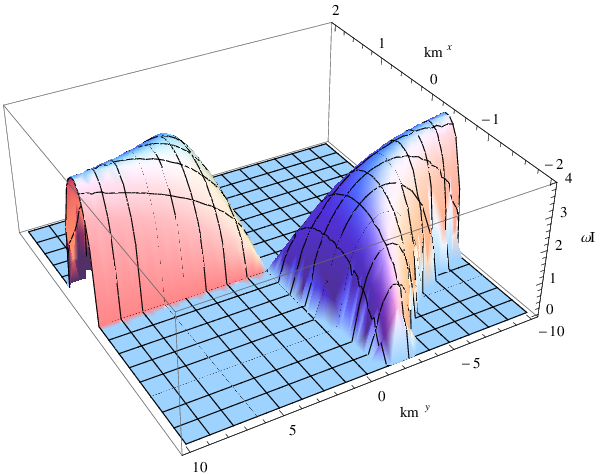}
}
\caption{ The growth rate $\omega_{I}$ of the MI given
by (\re{growthrate}), arising in our 2DNLS equation, exhibiting how it
changes (for $A_0=1.0, k^x=1.0 $) along the longitudinal $(k_m^x)$ and transverse ($k_m^y$)
directions, showing a strong directional preference.  } \label{fig:3}
\end{figure}
 The stability plot is drawn in Fig.  3 in the $(k^x_m,k^y_m)$ plane 
with the shaded region
showing the domain of MI.
\begin{figure}[!h]
\centering
{
 \includegraphics[width=4cm, angle=0]{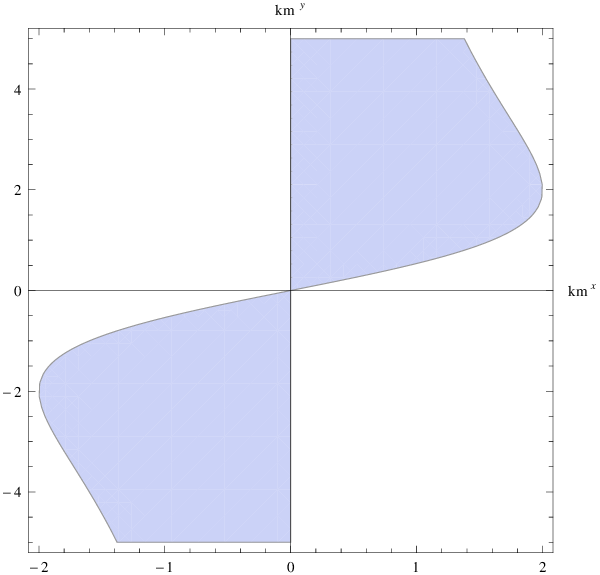}
}
\caption{ Graphical representation of the MI region,
where the instability can occur only within the shaded area (for fixed
values of $A_0=1.0, K^x=1.0 $).  The
instability region, showing dependence
 on the wave vector $(k_{x}^m ,k_{y}^m
)$, varies asymmetrically along the longitudinal and transverse direction,
as seen clearly from the figure.  } \label{fig:2} \end{figure}

  Both these figures show
clearly, that the behavior of MI as well as the growth
rate has a strong directional preference and range as observed also  
earlier in 2D models
 \cite{{prl2DWexp09,pla2Dnls00,{prl2Dnls10}}}.  We have confirmed such
properties through exact analytic result showing explicitly that in the
MI as well as in the growth rate the components
$(k_m^x,k_m^y) $ of the wave vector do not enter symmetrically, in addition
 with a
directional range $|{k_m^x}|<2|{k^x}|$.

A comparison here with the analysis of MI in case of
the known 1D
NLS equation \cite{agarwal} may be illuminating. 
 The condition for the onset of instability in the 1D case involves only the
nonlinear amplitude $ A_0$ expressed in the form $|{k_m^x}|<2A_0, $ while in
the present situation the condition is more complicated involving all
components ${k_m^x},{k_m^y}, {k^x}$ apart from $A_0 $, together with an
allowed
 range
 on the wave vector component, as
found above analytically and shown graphically in Fig.  3.  Similar
situation is  also true  for the growth rates, where in the 1D case it is
given by $\omega_I = \ | k^x_m|[(2A_0)^2-(k^x_m)^2]^{\frac 1 2}$ 
\cite{agarwal}, while in the
present case the form of $\omega_I $ is more complicated and depends on 
both longitudinal and transverse directions, as shown above.

Thus the overall picture for the onset of the 
 MI is   similar to that occurring in the 1D NLS
equation \cite{agarwal}, though in the  case of the 2D NLS equation
(\re{2dsInls}) the details are different and more intricate 
with a directional preference and range,  as seen also 
 for the  MI, initiating RW formation in some other
systems in higher space-dimensions {\cite{prl2DWexp09,prl2Dnls10,
prlCav09,pla2Dnls00}}.  We emphasize however, that in place of approximate
numerical result obtained earlier, we found here similar properties in exact
analytic form in our model.  This is a strong point of our exact model.
  As in the case of the
well studied 1D NLS model, we may expect 
 the MI to play  a key role  in the creation of 
 RWs based on  our 2D NLS model (\re{2dsInls}).
  

\section{3. Modelling of 2D rogue waves }

Apart from finding a novel 2D integrable equation  (\re{2dsInls}), 
 our aim, relevant to the
 present problem,  is to construct a 2D RW model
as  an exact solution of this equation. 
\subsection{3.1. Static lump soliton}

 Before presenting the
dynamical lump solution related to (\re{2dsInls}) we consider first its
static 2D lump-like structure: 
 \bea  
q_{P(2d)}(x,y)= e^{4iy}(u+iv), 
 \ u=G-1, \ v=-4yG, && \nonumber \\
 \mbox{ where} 
 \  G \equiv\frac 1 {F(x,y)},     \ F(x,y)=\alpha x^2+4y^2+ c, &&
\ll{2dPbr} \eea
  localized in both space directions and describing a fully developed RW. 
 One can check by direct insertion that (\re{2dPbr}), having two arbitrary
parameters
 $\alpha$ \  and $\ c $, is an
exact static solution of the 2D nonlinear equation 
(\re{2dsInls}). Solution (\re{2dPbr}), in spite of its close resemblance
 with  the well known PB 
solution (\re{Pbr}),  marks some important 
differences.
 The static  wave profile $|q_{P}(x,0)|$ (\re{Pbr0}),
  obtained from  PB solution (\re{Pbr})  (see Fig. 1)
 at time $t=0$
 is a curve, representing 
  full blown 1D RWs   
  admitting  no free parameters of relevance.  On the other hand 
\bea
 |q_{P(2D)}(x,y)| =({u^2+v^2})^{\frac 1 2}
=[(G-1)^2 + (4yG)^2]^{\frac 1 2},\label{modRog}\eea
 obtained from the static solution (\re{2dPbr}) represents
 a 2D lump
(see Fig. 4)
with two independent free parameters, significance of which will be
is explained below and shown in Fig. 4(a-d).

\noindent {\bf a.} 
\includegraphics[width=6cm, angle=-90]{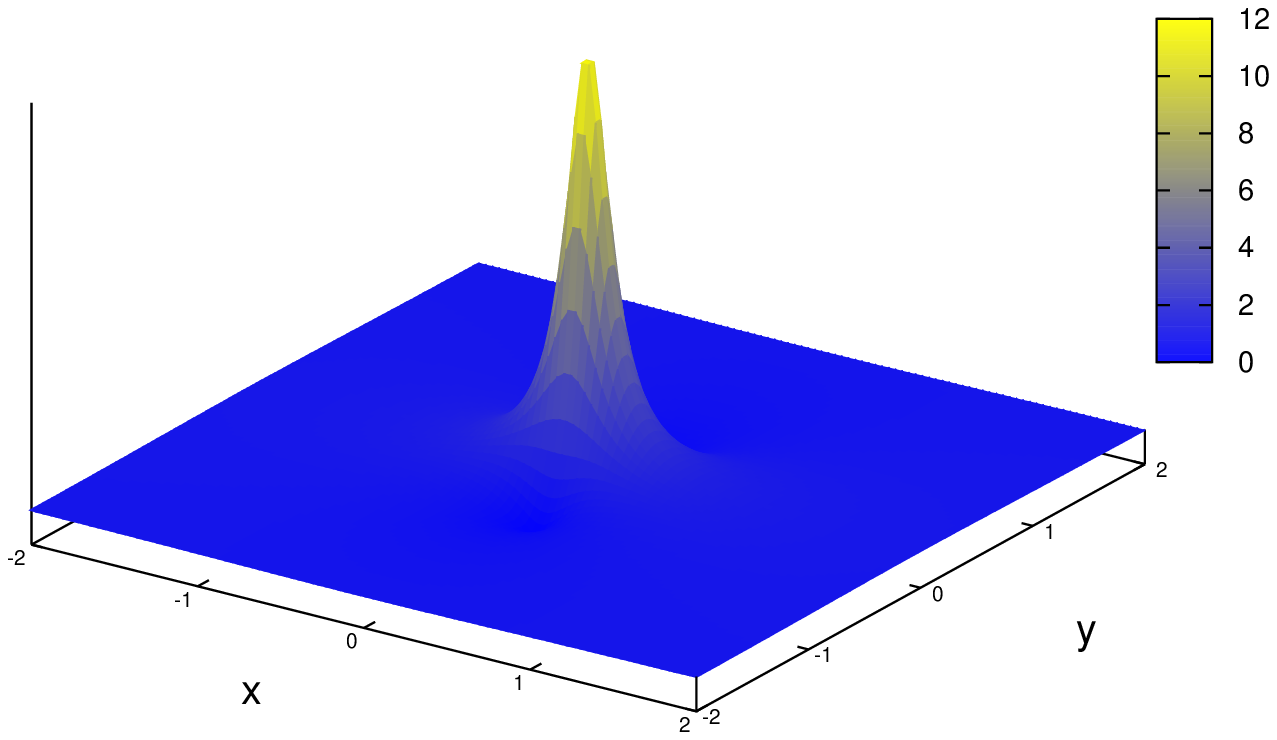}
\noindent{\bf b.}
{
\includegraphics[width=6cm, angle=-90]{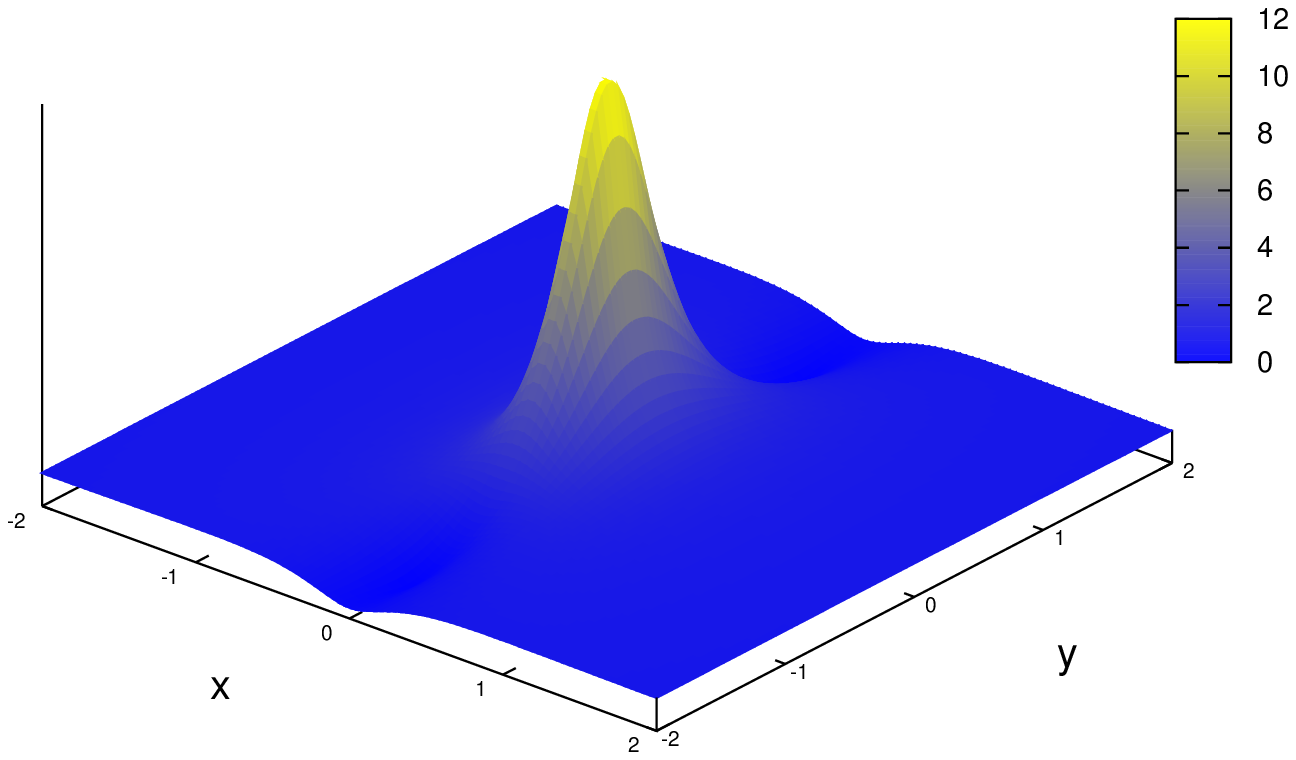}  
}
\\
\noindent {\bf c.}
 \includegraphics[width=6cm, angle=-90]{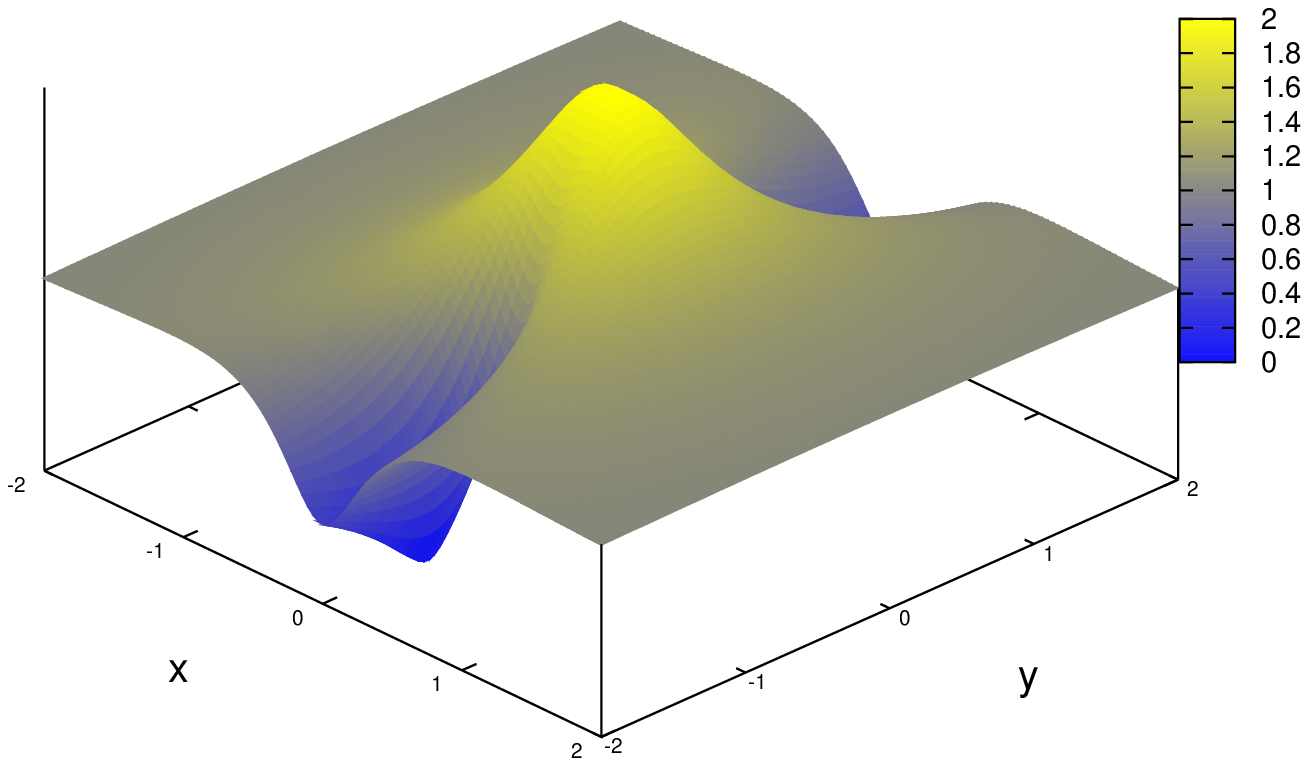}
\noindent {\bf d.}
 \includegraphics[width=6cm, angle=-90]{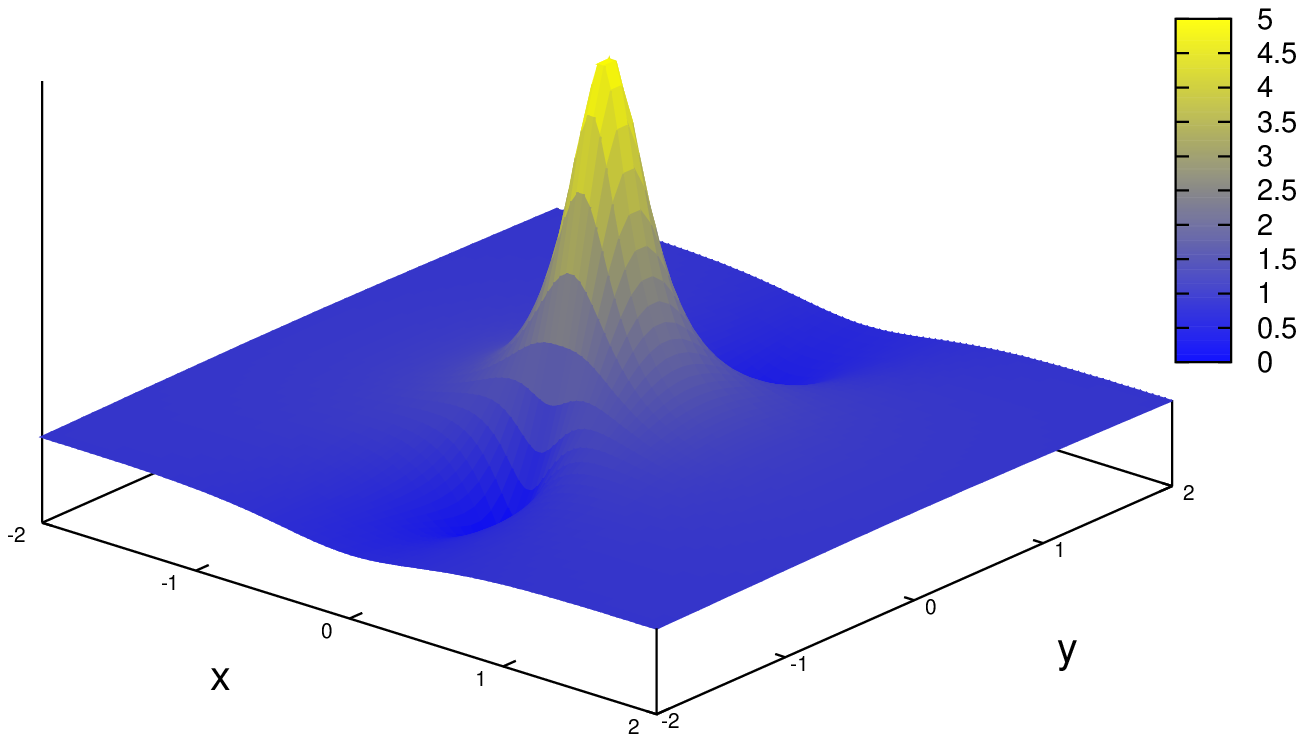}

{FIG 4. \ 
\small
 Full grown 2D rogue wave   modelled by   the modulus $(|q_{P(2D)}|)$
of the  static lump-solution 
(\ref{2dPbr}) with   different  shapes and sizes, 
  generated from the 
same single-peak solution. The  maximum amplitude and 
 modular inclination are tunable through    two free parameters $c$ and
$\alpha$ showing: (a) High amplitude: $12$ and high modular inclination ,
for $c=1/13,\ \alpha=4.0.  $ (b) High amplitude: $ 12$ but low modular
inclination, for $c=1/13,\ \alpha=0.4 $. (c) Low amplitude: $ 2$ and low
modular inclination, for $c=1/3,\ \alpha=0.4.  $ (d) Moderate amplitude: $
5$ and moderate modular inclination, for $c=1/6,\ \alpha=1.2 .$ The last
situation is the same as Fig.  5(c), obtained at $t=0$.  }

\subsection{3.2. Rogue wave with adjustable  amplitude, inclination and hole
waves} Note, that the static lump soliton (\re{2dPbt}), can be obtained from
the dynamical RW solution  (\re {Pbr0}) at the static point $t=0$ (Fig.  5c)
similar to static PB profile obtained from (\re{Pbmod}), and
  hence it  physically represents a full grown RW solution as shown in 
Fig. 4 (d).
Looking more closely into solution (\re{2dPbr}),
  for understanding the physical relevance  of its free parameters 
$c$ and $\alpha$,   we notice   that 
   the wave attains its maximum  amplitude: 
$ |q_{P(2D)}(0,0)| \equiv A_{rog}(c)=(\frac{1}{c}-1) $, 
at the center ($x =0 , \  y =0 $),
while 
at large
distances ($|x| \to  \infty , \  |y| \to  \infty $) the  wave 
 goes to   the background plane wave,  with its amplitude decreasing   to 
$A_{\infty}=1 $.
 Therefore the  maximum  amplitude 
 reachable by our RW solution, relative to that of the background wave
 is
$\frac {A_{rog}(c)} {A_{\infty}}=(\frac 1 c-1) $.
 Consequently, the amplitude of the full grown RW described by the lump
soliton can be changed continuously by changing parameter $c$
 (with $A_{rogue}(c)$ increasing with decreasing c) and could
therefore
be adjusted to fit the heights of any  observed RW. 
Consequently, the maximum RW amplitude 
 in our model can be made as high as desired,
 by decreasing the value of an arbitrary
    smooth parameter $c$ (see
Fig.  4 (a-d), for particular examples).  Comparing this
 situation with the
conventional 1D RW model given by the PB 
 (\re{Pbr}) and its generalizations \cite{MrogPRE09},
 as  mentioned above, we conclude that,   in   the 
well known class of  PB solutions, the maximum amplitudes
 reachable by the 1D RW  are given by the fixed  discrete odd numbers $2j+1 $, with
$j=1,2,3, \ldots $ and can be obtained
 by going only to different higher  solutions involving  more and more
complicated higher order polynomials.  The higher rational solutions having
free parameters \cite{Dubard,RWtriple11} become
 parameterless for a single-peak RW solution
\cite{RWtriple11}.  On the other hand, in our 2D RW model the maximum
amplitude $A_{rog}(c)$ can be varied continuously and increased as required,
by tuning an arbitrary parameter $c$ in the same single-peak, first order
solution
  (\re{2dPbr})  or its dynamical extension
 (\re{2dPbt}) (as shown in Fig. 4 (a-d)), making the
 model suitable for RWs with a diverse range of heights,
anywhere in the range 17-30
 meters in calm sea \cite{rogRev}-\cite{prlCurr11}, as observed 
  in  deep sea 2D  RWs.  

Extending the  modular inclination in case of 1D: $S_{P}^x(x)$, as defined
in (\ref{PBincl})
 we get for  the full grown 
2D RW solution (\re{2dPbr}) the  modular inclination  as
\begin{equation}
S_{P(2D)}^x(x,y)= \frac{\partial}{\partial x} |q_{P_{(2D)}}(x,y)|,
\ \ S_{P(2D)}^y(x,y)= \frac{\partial}{\partial y} |q_{P_{(2D)}}(x,y)|
\end{equation}
Focussing on the inclination $S_{P(2D)}^x(x,0)$
as observed  at the middle  of the wave front, we notice, that it 
is linked also to  another  free parameter $\alpha $ and 
attains its maximum

\begin{equation}
S_{P(2D)max}^x(x_{m},0)=- 2 \alpha x_{m}G^2(x_{m},0)
\end{equation}
at $x_{m} = \frac{\sqrt{c}}{\sqrt{3 \alpha}}$
  with  function
 $G(x,y) $ as defined in   (\re{2dPbr}).  We see that the maximum
modular inclination of a full grown 2D RW in our model depends on both the parameters $c$ and $\alpha$ in an intricate
way and 
 can   be changed continuously by varying two arbitrary parameters to fit
varied situations (Fig.  4(a-d)).  Note that this inclination will be
influenced by the physical steepness of the wave, contributing from the wave
vector of the career wave.  We can identify another intriguing feature of
our solution,
by noting  that the amplitude of the wave (\ref{modRog})
 falls to its minimum: $A_{0}=0 $,
 at $y=0, \ x= \pm x_0 $ where $x_0 =
\sqrt{\frac 1 \alpha (1-c)} $, which depends again on two free parameters.
 This   
significant feature emerging from  our RW model, 
as will be demonstrated below in Fig.  5(a,b),  is 
 related to the {\it hole-wave} formation observed during ocean RWs
\cite{ZakhJETP05,Rogwiki}.
\subsection{3.3. Topological consideration}

Though the static lump-solution  (\re{2dPbr}) can describe the profile 
of a full grown 2D RW, for modelling an 
evolving   realistic RW,  we 
need  to find   a time-dependent solution,
 which would smoothly
go to its static form   (\re{2dPbr})  at the  moment $t=0$.
 Our next aim therefore 
is to    construct a  dynamical lump soliton
out of the static lump-solution,
 to create  a true picture of a 
RW which  can appear and disappear  fast with time. 
However, for constructing 
such a solution  we have to clarify first, whether it is possible 
in principle for our lump soliton to disappear  without a 
trace,
 i.e. whether
 the soliton is free from all 
  topological restrictions, which  otherwise would prevent such a vanishing. 
The reason for such suspicion is due an interesting lesson from topology
stating that, when a complex field $q(x,y) $ is defined on a 2D space with
non-vanishing boundary condition $|q| \to 1 $ at large distances, but having
vanishing values $q \to 0 $ close to the center, we can define a unit vector
$\hat \phi= \frac q {|q|} $ on an 1-sphere $S^1 $.
 However, this   vector $\hat \phi  =(\phi^1,\phi^2) 
 $  is 
well  defined only at  the space boundaries: $\partial {\rm R}^2
\sim S^1  $ (since $q=0 $ at  inner points),
  realizing a smooth  map: $S^1 \to S^1 $ with possible nontrivial 
 topological charge $Q=n $. This charge   with integer values
$n=0,1,2,\ldots$,
 labels the 
distinct homotopy classes and is  defined
 as the degree of the map, which unlike a N\"other charge 
is conserved irrespective of the
dynamics of the system.
Such a situation occurs for example in type II superconductors
with the charge linked to  the quantized  flux of vortices for
 the magnetic 
field ${\bf B} (x,y) $
{\cite{Parsa79}} 
: \be
{2 \pi} Q= \int d {\bf S} \cdot {\bf B}=
 \int_C {\bf dl} \cdot {\bf A},
  \ll{Q} \ee   {where} 
$ \ {\bf B}=  {\bf {\rm curl} \  A}= {\bf  \hat  z}
 (\partial_x\phi^1 \partial_y \phi^2-
 \partial_x\phi^2 \partial_y \phi^1).$
Notice that, our  complex field solution
 $q_{P(2d)}(x,y) $   
     possesses clearly  the features 
 of $\hat \phi $  discussed above, since (\re{2dPbr})  
   goes to
a constant modulation $-e^{4iy} $ at large distances and  vanishes    
at  points $(0,\pm x_0) $.
Note that, such a solution  related to a  sphere to sphere 
 map can not go to a trivial configuration 
, if it belongs 
to a  homotopy class with nontrivial  topological charge:
 $Q=n, n=1,2,3,... $, due to conservation of the charge,
 with the only exception for the  class with zero charge $Q=0 $.
Therefore, for confirming  the possible appearance/disappearance  
property of a   RW  for  
solution  
(\re{2dPbr}), we have to
 establish first that in spite of defining a nontrivial  topological map,
 it   belongs nevertheless 
 to the  sector with topological  charge: $Q=0 $,  i.e. 
 our lump soliton is  indeed shrinkable   to the {\it vacuum} solution.    
For this we calculate explicitly
 the topological charge (\re{Q}) associated with 
  (\re{2dPbr})  as
 \be
{2 \pi} Q=  \int_C {\bf dl} \cdot {\bf A}=
 \int (dx  { A}_x+ dy  { A}_x), 
, \ll{Q1} \ee  where 
$ \ A_a=
 \phi^1\partial_a \phi^2,  \ \phi_1= {\rm Re}\  q/|q|, \ \phi_2= {\rm Im } \  q/|q| 
, $
where the contour integral along $x $ and $y$ are taken along a closed square 
 at the boundaries of the plane.
Substituting explicit form of solution $q(x,y) $ from  
 (\re{2dPbr})
and arguing about the oddness and evenness of the integrand with respect
to $x,y $ or checking  directly by any analytic computational package
  one can show
that the related charge is indeed $Q=0 $ and therefore  the solution  
 belongs to the
trivial topological sector as  we wanted. The intriguing reason behind this
fact is that,  the  two holes
 appearing here have      opposite charges resulting to 
 their  combined charge being zero. 
\subsection{3.4. Construction of dynamical lump soliton}

For constructing  a dynamical extension of the 2D static lump
 soliton (\re{2dPbr})
 we realize  that, a sudden change of amplitude with time, 
 as necessary to mimic the 2D RW  behavior, might result to a 
 non-conservation of
energy. This  however  can not be described by an integrable equation  alone,
 since the integrability demands  a strict 
conservation  of all charges and
therefore our integrable equation   (\re{2dsInls}) needs certain 
modification for allowing the 
 appearing/disappearing nature of its lump-solution. 
On the other hand, 
 the importance of   ocean currents in the formation of 
 RWs
  is  documented and  repeatedly emphasized  
{\cite{prlCurr11,ZakhJETP05,Ruban}}, which however 
 is absent in  equation  (\re{2dsInls}). This    motivates us    to
solve both these problems in one go, by  modifying  
  equation (\re{2dsInls})
with the  inclusion of  the effect
of an  {\it ocean current}, as in 
{\cite{prlCurr11}},  by  adding a term  in the    form
$I=-iU_cq_x $. 
For obtaining an exact dynamical 
 RW solution to the modified 2D NLS equation, we choose the current 
  flowing    along  longitudinal directions and changing   
 with  time and  location  as $U_c(x,t)=\frac {\mu t} {\alpha x}$. 
Looking closely into the structure of this current
term for the RW solution (\re{2dPbt}):
\begin{equation}
 I(x,y,t) = i (\frac{\mu t}{a x})\frac{\partial}
{\partial x}[q_{P(2D)}(x,y,t)] = -2 \mu t (4y- i)G^2 \ e^{4iy},
\label{Ocurrent}
\end{equation}
with $G$ as defined in (\re{2dPbt}), it becomes 
apparent, that the currents
 would flow  to  the center  of  formation of the RW ($x=y=0 $)
  from both of the longitudinal  and the trasverse sides, though with a
directional preference,  with  their magnitude $| I(x,y,t) | $  increasing
  as they approach to the center, however  stopping  completely 
at the moment of the full surge   at $t=0 $.  The picture gets
 reversed  after the RW event with   currents  flowing back quickly,
 away from
the center with the intensity of the current  $|I| $ 
diminishing as the distance
increases.
  Such an inflow and outflow of energy seems to be physically
consistent with the formation of a 2D ocean RW.
 Note that, though the current factor  $U_c $ looks ill-defined, the
multiplicative factor $q_x $ makes the 
 term  $I(x,y,t)$ well-behaved on the RW
solution (\re{2dPbt}), with the  ocean  current 
 term becoming a smooth and bounded
function in all space and time variables,
as  evident from    (\re{Ocurrent}).  It has been suspected in earlier
studies, that spatially nonuniform current should be responsible in the
development of ocean rogue waves \cite{Ruban}.  Such a nonuniform dependence
on space variables can be seen in our current term $I(x,y,t)$.
 Interestingly, 
the   modified 2D NLS equation ((\re{2dsInls}) with the inclusion of  
the  current term $I$)  admits now an exact   
  dynamical 2D extension of the Peregrine soliton in the analytic form,
 though the modified
equation loses its integrability in the sense, discussed in sect. 4.2. 
The dynamical RW solution   has a  
similar    form as
(\re{2dPbr}),   only with the      function $G $ becoming dynamical 
by the inclusion  of    time
variable :
\bea
q_{P(2d)}(x,y,t)= e^{4iy}[-1 + (1-i4y)G] ,
\nonumber \\
\mbox{}
 G \equiv \frac{1}{F(x,y,t)},
 F(x,y,t)= \alpha x^2 + 4y^2 + \mu  t^2+  c.
\ll{2dPbt} \eea
The  arbitrary parameter $\mu $ appearing in  the  solution   
 (\re{2dPbt}) is related to the ocean current and can  control how fast the RW would
appear and how long it would stay.  Note again that
  (\re{2dPbt}) at $t=0,$ representing a full grown RW  (see Fig 5c))
coinciding with  the exact static lump-solution  (\re{2dPbr})
 of the 2D NLS equation 
(\re{2dsInls}) (as in Fig. 4d)), justifying the
physical relevance of the  static lump solution.
At this stage, a comparison between 1D PB soliton 
\begin{equation}
 q_{P}(x,t) = [-1 + \frac{(1-4 i t)}{x^2+4 t^2 +\frac{1}{4}}] e^{-2 i t}
\end{equation}
and our 2D lump soliton 
\begin{equation}
 q_{P(2D)}(x,0,t) = [-1 + \frac{(1)}{\alpha x^2 + \mu t^2 + c}], \ \mbox {at}
\ y=0, \label{2Dy0}
\end{equation}
  might be interesting. This shows that though there is some similarity 
between these two solutions, there are many differences as well at $y=0 .$
In  the absence of the  transverse coordinate, the 2D solution (\re{2Dy0})
 of our modified
equation,  becomes real, though still having $3$ independent
 free parametrs. The 1D PB
soliton on the other hand is complex with a breathing mode, but without any
free parameter. 

We should mention here, that the 2D extension of  PB
solution (\re{2dPbt}), unlike the standard 1D PB, 
unfortunately could not be derived   as a limiting case from the 
 breather solution of  2D NLS equation (\re{2dsInls}),
 due to the  two-dimensional nature 
of the solution and has to be constructed by direct insertion 
through an ansatz. 
\subsection{3.5. Proposed 2D  rogue wave model and its dynamics}
 It is convincingly demonstrated in Fig. 5 (fixing the  free parameters
to certain values), how 
 the envelop wave $|q_{P(2D)}(x,y,t)|$ corresponding to the exact dynamical 
2D  lump-soliton (\ref{2dPbt}),
dependent on time t and two space variables $x$ and $y$ on a plane,
 evolves  from 
a background  plane wave
 existing in the distance past and how it could acquire a sudden 
  2D hole at the centre $(x=0,y=0)$ 
at the moment 
$t_h=-\sqrt{\frac 1 \mu (1-c)}$,  ($t_h=-0.83 $ for $c=1/6, \mu=1.2$ in
Fig. 5a), 
  as told in marine-lore
{\cite{Zakh09,pla2Dnls00,Rogwiki}}. The hole subsequently 
  splits into two and shift apart from the centre (Fig.  5b)), to make space
 for a high steep upsurge of the lump forming the full grown RW (Fig.  5c)
 at time $ t=0$.  Note that we have derived analytically
the exact positions of these holes in sect. 3.2. 
With the passage of time the picture gets reversed and
the 2D RW disappears fast  into the background waves with the 2D  holes
merging at the centre  and vanishing again. Thus
our model describes  vividly well the reported picture of the ocean 
 surface RWs 
{\cite{prlCurr11,prlmulti11,ZakhJETP05}} 
 as well as
those found in  large scale 2D experiments
{\cite{prl2DWexp09}}. Since our model is an exact one , we could work out these details analytically.
The surface RWs modelled by  our  solution (\re{2dPbt}) and
 as visible  from Fig. 5 (similarly from 
 solution (\re{2dPbr})  and   Fig. 4),
shows a distinct directional
preference  and   an  asymmetry between the
 two space variables $x,y,$ (similar to the report of
 \cite{prlCurr11,prlmulti11}).
 The  maximum amplitude attained by the full
grown  RW  (as shown in Fig 5c) is
five-times that of the background waves, due to our choice  $c=1/6$.
Examples of  other 
  amplitudes and modular inclination of  full grown RWs for  some other choices 
of the free parameters $ c$ and $\alpha$, as   modelled by the 
static solution
(\re{2dPbr}), are presented already in Fig. 4 (a-d).




{\bf a.}\includegraphics[width=6.cm, angle=-90]{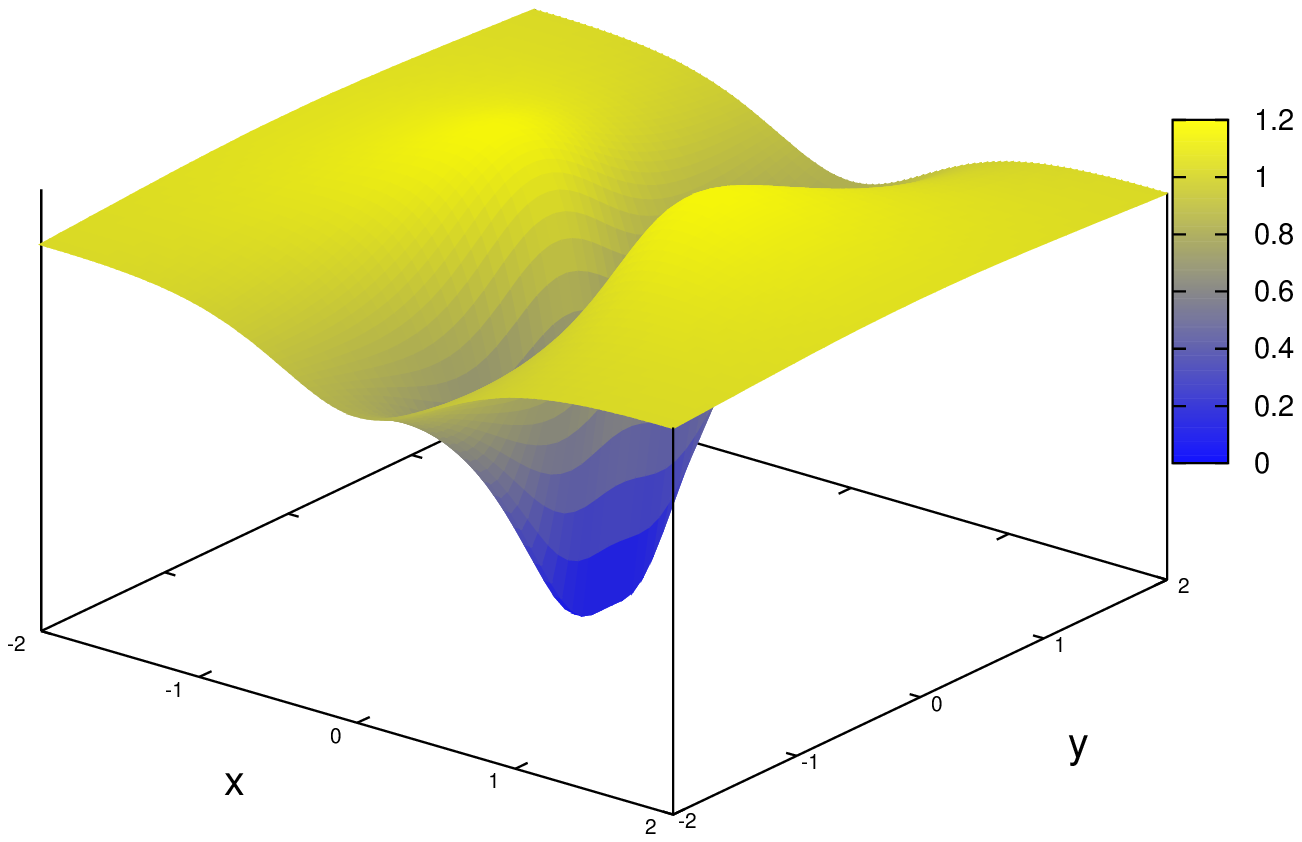}
{\bf b.} 
\includegraphics[width=6.cm, angle=-90]{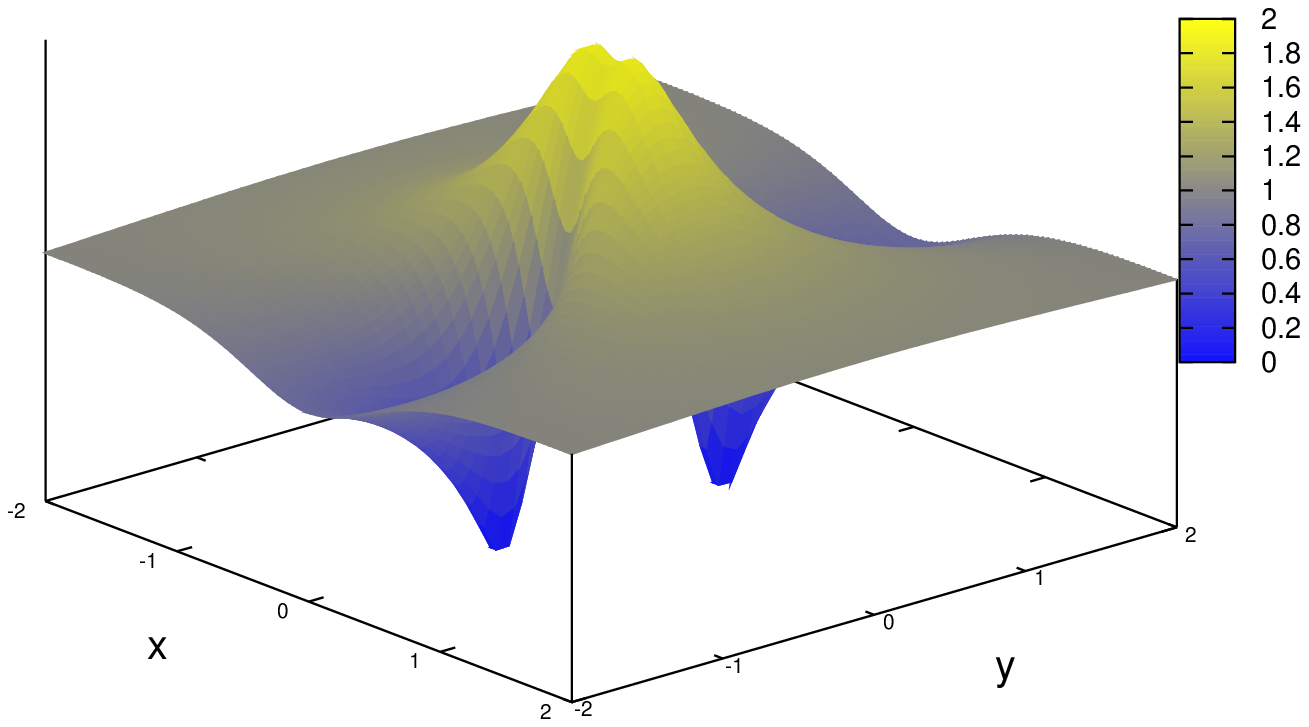}  
{\bf c.} \includegraphics[width=6.cm, angle=-90]{a12c033.eps}

{FIG 5. \
\small
 Snap shots  of  
a 2D rogue wave with  2D holes
   during its  formation at different time, described  by the  modulus
 $|q_{P(2D)}(x,y,t)|$ of 
 the dynamical lump-soliton 
 (\re{2dPbt}) with 
  parameter values  $c=1/6,\  \alpha=1.2 $ and $ \  \mu=1.2$, at three
crucial  moments of time: (a) At  $t=t_h=-0.83$: Creation of a 2D hole at
the centre.  (b) At  $t=-0.40 $: The hole splits into two are drifting
away from the centre. (c) At  $t=0.0$: The full grown RW,
 corresponds also 
to the static lump solution
 (\re{2dPbr}), as shown in Fig. 4(d).
}

\section{4. Physical origin of the proposed 2D NLS equation and its
integrability} We have shown that the 2D NLS equation (\re{2dsInls})
which is equivalent to nonlinear equation (\re{2dInls})
 can give an exact model, considerably
successful in describing realistic 2D rogue waves.  In this section we show
the direct link of the  2D NLS equation with basic hydrodynamic equations. 
Moreover we show the underlying integrable structures of the proposed equation.

\subsection{4.1. Derivation of the integrable 
2D NLS equation from  basic hydrodynamic  equations}

For  emphasizing the physical   significance  of our main nonlinear
integrable equation (\re{2dsInls}), on which the ocean rogue wave model is
based, we show its direct link with
 basic hydrodynamic  equations. 
The procedure is based on a multi-scale expansion, at par with the
celebrated equations like KdV, NLS etc   \cite{BOOKMP,Ablowitz},
 though  one should  include here  an extra space dimension   with an
 asymmetric scaling in space variables, considering the perturbative
 expansion to the next higher order.  This is consistent however with the
 modelling of an ocean rogue wave, which is a surface phenomena with a
likely broken space symmetry and directional preference \cite{prlCav09}.
 Before entering into the detailed
calculation, three dimensionless entities:
 $\epsilon = \frac{a}{h_{0}}$ , $\delta = \frac{h_{0}}{\lambda_{x}}$ and
$\mu = \frac{\lambda_{x}}{\lambda_{y}} $ are defined where $a$
 is the maximum
amplitude, $h_{0}$ is the constant water depth, $\lambda_{x}$ and
$\lambda_{y}$ are the
 wavelengths of the surface wave along longitudinal and transverse
directions.  The nonlinear parameter $\epsilon$ is responsible for  the
slow evolution of a harmonic wave of wavenumber {$k_{x}$, $k_{y}$.  The wave
is thus slowly modulated as $\epsilon$ tends to $0$ and 
 therefore this small parameter can be used for perturbative expansion. 
Smallness of $\epsilon$ is consistent with the deep water limit with $a<<
h_{0}$ and hence  with the formation of oceanic rogue waves.
Note that parameters $\epsilon $ and   $\delta$
are similar to those appearing in the derivation of the well known 1D NLS
equation, with $\epsilon $ small  and   $\delta$ without any restriction, as
also true in our case. However, an additional parameter
$\mu $, also without any restriction on its value appears in our 2D case,
due to the presence of an additional transverse direction.

 The first step in the derivation  is to write the basic hydrodynamic
 equations for inviscid, irrotational and incompressible fluid in
 dimensionless variables,
for the velocity potential field
 $\phi(t,x,y) $ and the gravity  wave $\eta(t,x,y) $ as the free surface
displacement above the mean water depth $h_{0}$ in the form \begin{equation}
 \phi_{zz} + \delta^2 (\phi_{xx} + \mu^2 \phi_{yy}) = 0,\label{LapE}
\end{equation}
at $0 < z < 1+ \epsilon \eta$, which  comes from continuity equation.
The equation
\begin{equation}
 \phi_{z} = \delta^2 [\eta_{t} + 
\epsilon(\phi_{x}\eta_{x} +\mu^2 \phi_{y}\eta_{y})]\label{nonbc1} ,
\end{equation}
  called kinematic condition,
 is valid on z = 1 + $\epsilon \eta$. The equation
\begin{equation}
\phi_{t} + \eta + \frac{1}{2}\epsilon 
[\frac{\phi^2_{z}}{\delta^2} + \phi^2_{x}+ 
\mu^2 \phi^2_{y}] = 0 \label{nonbc2}
\end{equation}
 is the Bernoulli's equation and also valid at z = 1 + $\epsilon \eta$, while
\begin{equation}
 \phi_{z}  = 0 \label{fbc}
\end{equation}
is the fixed boundary condition valid at z = 0, i.e. at the bottom.

We introduce new variables with different scaling through $\epsilon$ as 
\begin{equation}
 \xi = k_{x}x + k_{y}y -\omega t, \
 \zeta =\epsilon( x - M_{x}t), \
 Y = \epsilon^2 y, \
 \tau = \epsilon^3 t,
\end{equation}
where $\omega$, $M_{x}$ are
 frequency and velocity parameters to be determined later.
  Note, that  the two space variables are treated 
with a non-symmetric scaling and
 using these set of variables, equations (\re{LapE} -\re{fbc}) become 
\begin{eqnarray}
 \phi_{zz} + \delta^2(k_{x}^2 \phi_{\xi \xi} + \epsilon^2 
\phi_{\zeta \zeta}+ 2 \epsilon k_{x}\phi_{\zeta \xi})
+\mu^2 \delta^2 (k_{y}^2 \phi_{\xi \xi}
 + \epsilon^4 \phi_{YY}+ 2 \epsilon^2 k_{y}\phi_{Y \xi}) =0  \label{mLapE}
 \end{eqnarray}
\begin{eqnarray}
 \phi_{z} = \delta^2
 [-\omega \eta_{\xi} - \epsilon M_{x}\eta_{\zeta}
 +\epsilon^3 \eta_{\tau}] +
\epsilon \delta^2(k_{x}\phi_{\xi}+\epsilon \phi_{\zeta})
(k_{x}\eta_{\xi}+
\epsilon \eta_{\zeta})+
\mu^2 \epsilon \delta^2(k_{y}\phi_{\xi} 
+\epsilon^2 \phi_{Y})(k_{y}\eta_{\xi}+\epsilon^2\eta_{Y}) \label{mnonbc1}
\end{eqnarray} and
 \begin{eqnarray}
 [-\omega \phi_{\xi} - \epsilon M_{x}\phi_{\zeta} +\epsilon^3 \phi_{\tau}] + \eta +
(\frac{\epsilon}{2 \delta^2}) \phi_{z}^2
 + \frac{\epsilon}{2}(k_{x}\phi_{\xi}+\epsilon \phi_{\zeta})^2+
\frac{\mu^2 \epsilon}{2}(k_{y}\phi_{\xi} 
+\epsilon^2 \phi_{Y})^2 = 0 \label{mnonbc2}
 \end{eqnarray}
 both valid at $z = 1 +  \epsilon \eta$, while
\begin{equation}
\phi_{z}= 0, \ \mbox{at} \  z=0. \label{mfbc}
\end{equation}

Seeking asymptotic solution of these equations in the series form 
\begin{eqnarray}
 \phi = \sum _{n=0}^{\infty}  \epsilon^n \phi_n(\xi,\zeta,Y,\tau,z) , \
  \ \eta=\sum _{n=0}^{\infty} \epsilon^n \eta_n(\xi,\zeta,Y,\tau)
  \end{eqnarray} 
and using this expansion of dependent variables along with
the scaled independent variables, different sets of equations are obtained
from the basic set (\re {mLapE})-(\re {mfbc}) at different powers of
$\epsilon$.  In each $\epsilon$ order different equations are obtained for
various powers of $E$ and $E^*$.  We would consider these equations
sequentially at each order of parameter $\epsilon$.

{\bf 1) $\epsilon^0 $ order :}
The solution of interest in this case takes the form 
\begin{eqnarray}
 \phi_{0} = f_{0} + F_{0}E + F_{0}^*E^* ,  \eta = A_{0}E + A_{0}^*E^*,
\end{eqnarray} where
$  F_{0}(\zeta,Y,\tau,z)$ 
 , $  A_{0}(\zeta,Y, \tau)$ are complex functions 
with $F_{0}^*$, $A_{0}^*$ as complex conjugates, 
while $f_{0}(\zeta,Y, \tau)$ is a real function and $E = exp{(i \xi)}$.

Using (\re{mLapE}) and (\re{mfbc}), $F_{0},$ can be determined as
\begin{equation}
 F_{0}= G_{0} \cosh{(\delta K_{1} z)},\
\mbox{where} \ G_{0}= \frac{-i A_{0}\omega \delta}{K_{1}\sinh{(\delta K_{1})}}
 ,
 K_{1}=\sqrt{k_{x}^2+\mu^2 k_{y}^2}. 
\end{equation}
Using other two nonlinear  boundary conditions (\re{mnonbc1}),
 (\re{mnonbc2}) we obtain the dispersion relation \
 $\omega^2 = \frac{K_{1}}{\delta}\tanh{(\delta K_{1})}$

{\bf 2) $\epsilon$ order :}
Expanding $\phi_{n}$, $\eta_{n}$ as 
\begin{eqnarray}
 \phi_{n} = \sum _{m=0}^{n+1} F_{nm} E^m +c.c , \
  \ \eta_{n} = \sum _{m=0}^{n+1} A_{nm} E^m +c.c ,
  \end{eqnarray}
where $ F_{nm}(\zeta,Y,\tau,z)$  
and $ A_{nm}(\zeta,Y, \tau)$ are to be determined for various 
powers of $ E$, at each powers of $\epsilon$ .

At $\epsilon$ order ,the components
$F_{10},F_{11},F_{12},A_{10},A_{11}, A_{12}$  and the velocity parameter $M_{x}$
are determined from the equations corresponding to $E$, $E^2$ and $E^0$,
explicit forms of which are appended in $A11$.

{\bf 3) $\epsilon^2$ order:}
At this order  a  NLS type equation (Space coordinate Y replacing the time coordinate) is
obtained, collecting the coefficients of $E$ from (\re{mnonbc1}), (\re{mnonbc2}) and by using the quantities, already determined.
Before calculating the final form of this equation some other components namely $F_{21}, F_{20}, f_{0\zeta}$
 at this order need to be evaluated, which are given in the appendix $A22$. 

The final form of the NLS like equation is obtained eliminating the unknown terms
and expressing other terms through the single function $A_{0}$ as
\begin{equation}
i \alpha_{1} A_{0Y} + \alpha_{2} A_{0 \zeta \zeta} + \beta_{2} |A_{0}|^2 A_{0} = 0,\label{Spacenls}
\end{equation}
where the constant coefficients $\alpha_{1}$, $\alpha_{2}$ and $\beta_{2}$ are also given in appendix $A22$

Following the same procedure the components
 $F_{22}$,$A_{22},A_{20}$,$f_{0Y}$ are determined
 ,which 
we are not furnishing here due to their cumbersome expressions.

{\bf 4) $\epsilon^3$ order:}
In this order an evolution equation is obtained, for which 
some relevant components i.e. $F_{31}$,$F_{30}$ etc,
 are also determined by continuing with the same 
procedure. The explicit forms of these coefficients presented in $A33$.

The evolution equation  obtained by 
using equation (\re{mnonbc1}), (\re{mnonbc2}) and collecting coefficients of $ E$
takes the form
\begin{eqnarray}
 i a A_{0\tau}+ \alpha_{31} A_{0\zeta Y} + i  
\bar{\beta}_{32} A_{0}^2 A^*_{0 \zeta}+ i \bar{\beta}_{31}
 |A_{0}|^2 A_{0 \zeta}+
i e G^*_{11} A^2_{0} + 
i f G_{11} |A_{0}|^2 + i \alpha_{32} A_{0\zeta\zeta\zeta} = 0, \label{2devolE}
\end{eqnarray}
 where $a, \alpha_{31},\bar{\beta}_{32} , \bar{\beta}_{31}
,e ,f, \alpha_{32} $ are real constants dependent on parameters $k_{x}, k_{y},
 \mu, \delta$.

 If it is assumed, that the term $G_{11}$ 
 depends also on $A_{0}$ like the other terms as 
$F_{0} \sim A_{0}$ and $G_{12}, A_{12} \sim A_{0}^2$ etc. (see Appendix) ,
then the only consistent relation would be  $G_{11} = P_{1} A_{0 \zeta}$, where $P_{1}$ is
a real constant, dependent only on $k_{x}, k_{y}, 
\mu, \delta$. Using this relation  in (\re{2devolE}) one simplifies
 it in the form
\begin{eqnarray}
 i a A_{0\tau} + \alpha_{31} A_{0\zeta Y}+
i\alpha_{32} A_{0 \zeta \zeta \zeta}
+ i(\beta_{31} |A_0 |^2
A_{0\zeta} + \beta_{32}A_0 ^2 A^*_{0\zeta}) = 0,
 \label{2dnlsNI}
 \end{eqnarray}
where  $  
 \beta_{31}, \beta_{32},$ are another set of constant coefficients expressed
through earlier coefficients. Notice that 
the above equation (\re{2dnlsNI}) is similar to but not the 
same as our integrable 2D NLS
equation due to the appearance of the term 
$i\alpha_{32} A_{0\zeta\zeta\zeta} $. However fortunately we have another
equation (\re{Spacenls}) at our disposal, obtained at a lower order.
Taking derivative of (\re{Spacenls})  with respect to
 $\zeta $ we derive the relation
\begin{equation}
 i \alpha_{2} A_{0\zeta \zeta \zeta} = \alpha_{1} A_{0 \zeta Y}-i \beta_{2}(|A_{0}|^2 A_{0})_{\zeta}
\end{equation}
using which we can eliminate this
 unwanted term from (\re{2dnlsNI}) to obtain an equation in the form
\begin{equation}
 i C_{0} A_{0\tau} + C_{1} A_{0\zeta Y} +
  i C_{2} A_{0}(A_{0} A^*_{0\zeta} - A^*_{0} A_{0\zeta}) = 0,\label{final}
\end{equation}
under the condition on the coefficients of the original equation as
\begin{equation}
 \frac{\beta_{2}}{\alpha_{2}} 
= \frac{(\beta_{32}+ \beta_{31})}{3 \alpha_{32}}
\label{constrain}\end{equation}
Rescaling $\zeta$, $Y$ and $\tau$ and renaming $A_{0}$ 
equation (\re{final}) goes directly to the 2D NLS equation 
(\re{2dsInls}), which is equivalent to 
(\re{2dInls}) proposed by us.
Note that  constraint (\ref{constrain}), we have to impose  for
deriving our integrable 2D NLS equation from the basic hydrodynamic
equations, though
does not hold for general  water wave problems,
 this loss of generality is compensated for by the
  gain of our important exact results. This  in general is 
  true for   all integrable models.

\subsection {4.2. Integrable structures of the proposed equation:}
 We present here the associated integrability properties of equation (9).
 The one-soliton solution of this equation is given in the form  
 \  $ q_{s(2d)}(x,y,t)= {\rm sech}\kappa(y+\rho x-vt)
 e^{i(k_1x+k_2y+\omega t)},\ $
while allowing also higher
 soliton solutions and infinite set of conserved quantities \cite{arXiv12}.
 One can also find
 the associated linear  system
 $$\Phi_y=U(\la)\Phi, \Phi_t=V(\la)\Phi, $$  with a Lax pair given by
\bea U(\la )\equiv V_2(\la)=2\la V_1(\la)+V_2^{(0)},\
V(\la)\equiv V_3(\la)=2\la V_2(\la)+V_3^{(0)}\ll{V23}\eea
where 
\bea
V_1(\la)=i(\la \sigma^3+U^{(0)}), \
V_2^{(0)}=\sigma^3(U^{(0)}_x-i{U^{(0)}}^2)\nonumber\\
V_3^{(0)}=
-\sigma^3U^{(0)}_y -[U^{(0)},U^{(0)}_x], \
U^{(0)}= q \sigma^+ + q^* \sigma^-
, \ll{Lax}\eea
with $\sigma^a, a = \pm,3, $ Pauli matrices,
 the flatness condition: $U_t-V_y+[U,V]=0$, of which generates
 our 2D NLS equation
(\re{2dsInls}). 
Note that unlike the  known Lax pair of the 1D NLS, the pair
 $U(\la),V(\la)$ associated to our system
 have higher order dependence on the spectral
parameter $\la $.
  It is not difficult to show,
 that the flatness condition yields from (\ref{Lax})
 different relations at different powers of $\lambda$. The equation
 linked  to the   $\lambda$  
 corresponds to our $(2+1)$-dimensional
 NLS equation (\re{2dsInls}),
while the relation with  $\lambda^0$  gives
  another intriguing 
  nonlinear  equation
 \be  i  q_{xt}+q_{ y  y}  +2i|q|^2q_y+2q_x(qq^*_{ x}-q^*q_{ x})=0. \ll{2dsInlsE} \ee
Our main concern here however is the  2D NLS equation (\re{2dsInls}), 
 which we intend to use for constructing a 2D rogue wave model.
Note however, the modification of (\re{2dsInls}) by the addition of the
current term as considered in the sect. 3.4, though yields exact analytic RW
solution no longer remains integrable in the sense described here.

\section{5. Concluding remarks}
We conclude by listing a 
 few distinguishing    features of our proposed  dynamical lump  soliton
 (\re{2dPbt}),
which are  important    for a realistic  ocean RW model.

1) This is the first 2D dynamical RW model given in an analytic form.

2) It is a 2D extension of   Peregrine like soliton, representing an exact 
 lump solution linked to a novel $(2+1) $-dimensional
 integrable NLS equation,
derivable from the basic hydrodynamic equations.

 3) The dynamics of the RW solution is induced by a ocean current term and
controlled by it. Importance of the current in the formation of RW is
 strongly emphasized   \cite{prl2DWexp09,prlCurr11}, though perhaps 
for the first time this effect in 2D is attempted to be analyzed analytically
in our model. 

 4)
 Both the height and the inclination of the single peak RW
are  adjustable  by two independent free parameters present of our model.

 5) The fastness of   appearance of the RW    
and the  duration of  its stay can be  regulated  by yet another 
parameter linked to   the ocean current.

6) The proposed solution and MI exhibit   
  broken spatial symmetry  as well as 
a directional preference,
  which are suspected to be the crucial features in the formation of a 2D RW \cite{prl2DWexp09,prlCav09,prl2Dnls10,pla2Dnls00}.
Note again that these features  obtained earlier through observation or numerical simulation, found and confirmed in our model through
exact analytic result.

7) Strange appearance (and disappearance) of a  2D hole just before (and after)  
 the formation of the rogue
wave
{\cite{pla2Dnls00,ZakhJETP05,Rogwiki}}
  is also confirmed in our  model, graphically as well as by analytic findings.

 In comparison the original Peregrine soliton (\ref{Pbr}) (together with its
higher order solutions),
 by far  the
 most popular   model of the rogue wave,  does not
 exhibit most  of these essential properties, due to its inherently
 one-dimensional
 nature and absence of free parameters.  Therefore, while the class of
Peregrine solitons are successful in modelling 1D rogue wave
like structures observed in many experiments, the
 two-dimensional rogue wave model reported here 
should complement it, to stand close to a
 realistic model for ocean surface rogue waves.

We hope that, this breakthrough in describing large ocean 
 RWs by an   analytic dynamical lump-soliton with adjustable height,
 inclination and duration would also be valuable for
experimental findings of two-dimensional 
 RWs in other systems like capillary fluid waves
{\cite{prlcapil10}} 
  optical cavity  waves
{\cite{prlCav09}}
   and basin water waves 
{\cite{prl2DWexp09}}. Derivation of our exact lump soliton
from the breather solution of the  integrable 2D 
 NLS equation presented here, in a  systematic way
 as well as to find  higher order rational lump
solutions would be challenging theoretical problems.   

\section{Acknowledgement}
A.K. thanks P. Mitra for valuable discussions and the AvH Foundation
for  support.


\section {Appendix:}
\noindent {\bf A11: Coefficients appearing in order $\epsilon$ :}

$
 F_{10} = G_{10}(\zeta,Y,\tau),
$
$
A_{10} = M_{x}f_{0\zeta} - 
\frac{2 \delta K_{1}}{\sinh{(2 \delta K_{1})}}|A_{0}|^2 
$

$
 F_{12} = G_{12}\cosh{(2 \delta K_{1}z)},
$
  where  $G_{12}= \frac{-3 i \omega \delta^2}{4 \sinh^4{(\delta K_{1})}} A_{0}^2$,

$A_{12}= \frac{\delta K_{1}\cosh{(\delta K_{1})}}{2[\sinh{(\delta K_{1})}]^3}
[1 + 2 \cosh^2(\delta K_{1})]A_{0}^2$

$
F_{11} = G_{11}\cosh{(\delta K_{1}z)}-\frac{i \delta k_{x}}{K_{1}}G_{0 \zeta}z
 \sinh{( \delta K_{1}z)} 
$,

$A_{11}= i \omega [G_{11}\cosh{(\delta K_{1})}-\frac{i \delta k_{x}}{K_{1}}G_{0 \zeta}
 \sinh{( \delta K_{1})}]+ M_{x}[G_{0\zeta}\cosh{(\delta K_{1})}] $

The velocity parameter:
$ M_{x}= \frac{\omega k_{x}}{2 K_{1}^2}[1 + \frac{2 \delta K_{1}}{\sinh{(2 \delta K_{1})}}]$

\noindent {\bf  A22: Coefficients appearing in order $\epsilon^2$ :}

$
 F_{21} = G_{21}\cosh{(\delta K_{1}z)}-\frac{i \delta k_{x}}{K_{1}}G_{11 \zeta}z
 \sinh{( \delta K_{1}z)} -\frac{i \delta k_{y}}{K_{1}}\mu^2G_{0 Y}z
 \sinh{( \delta K_{1}z)} +
G_{0 \zeta \zeta}[(-\frac{\delta}{2 K_{1}})z \sinh{( \delta K_{1}z)} + 
(\frac{\delta k_{x}^2}{2 K_{1}^3}) z \sinh{( \delta K_{1}z)}-(\frac{\delta^2 k_{x}^2}{2 K_{1}^2})
z^2 \cosh{( \delta K_{1}z)}],
$

$
f_{0\zeta} = \frac{1}{(1- M_{x}^2)}[-\frac{2 M_{x}\delta K_{1}}
{\sinh(2 \delta K_{1})}-\frac{2 \omega \delta k_{x}\coth{(\delta K_{1})}}{K_{1}}]|A_{0}|^2,
$

$ F_{20} = -\delta^2 f_{0 \zeta \zeta}\frac{z^2}{2} + G_{10}(\zeta,Y,\tau)$,

$\alpha_{1} = -k_{y} \mu^2  \tanh{(K_{1}\delta})\frac{[2 K_{1}\delta +\sinh{(2 K_{1} \delta)}]}{2 \omega^3 \cosh^2{(K_{1} \delta)}}$,

$\alpha_{2} =\frac{\delta}{2 \omega K_{1}^3 }    
[K_{1}^3 \delta\{2 M_{x}^2 - \frac{1}{\cosh^2{(K_{1}\delta)}}\}-K_{1}^2 
\tanh{(K_{1}\delta)}+k_{x}^2 \tanh{(K_{1}\delta})
+ 4 K_{1}k_{x} M_{x}\delta^3 \omega^3- K_{1} k_{x}^2 \delta \{1+ \tanh^2{(K_{1}\delta})\}]     
$,

$\beta_{2} = - \frac{\delta^2 }{\omega (1- M_{x}^2)}
[4 k_{x}^2 +K_{1}
 \delta \frac{1}{\sinh{(2 K_{1}\delta)}}\{K_{1}^2((-1+ M_{x}^2)(8 + \cosh{(4 K_{1}\delta)}
 \frac{1}{\sinh^2{(K_{1}\delta})}+
 2 \frac{1}{\cosh^2{(K_{1}\delta})})
+ 8 k_{x} M_{x} \omega\}]
$

\noindent {\bf A33: Coefficients appearing in order $\epsilon^3$ :} 
\begin{equation}
 F_{30} = - \delta^2 F_{10\zeta \zeta} \frac{z^2}{2}+ G_{30}(\zeta,Y,\tau),
\end{equation}
$
 F_{31} = G_{31}\cosh{(\delta K_{1}z)} + G_{21\zeta}
 \{(\frac{-i \delta k_{x}}{K_{1}}) z \sinh{(\delta K_{1} z)}\}+
 G_{11\zeta\zeta}\{(\frac{i \delta k_{x}}{K_{1}})^2\frac{z^2}{2} \cosh{(\delta K_{1} z)}- 
(\frac{i \delta k_{x}}{K_{1}})^2 (\frac{z}{2 \delta K_{1}}) \sinh{(\delta K_{1} z)} - 
(\frac{\delta}{2 K_{1}})z \sinh{(\delta K_{1} z)}\} + 
(\frac{i \delta k_{x}}{K_{1}})G_{0\zeta\zeta\zeta}\{(\frac{\delta}{2 K_{1}})
(z^2/2)\cosh{(\delta K_{1}z)}- (\frac{\delta}{2 K_{1}})(\frac{z}{2 \delta K_{1}})\sinh{(\delta K_{1}z)} -
(\frac{\delta k_{x}^2}{2 K_{1}^3})\frac{z^2}{2} \cosh{(\delta K_{1} z)} + \frac{\delta k_{x}^2}{2 K_{1}^3}(z/2 \delta K_{1})
 \sinh{(\delta K_{1}z)}+ 
(\frac{\delta}{2 K_{1}}) \frac{z^2}{2}\cosh{(\delta K_{1}z)}-
 (\frac{\delta}{2 K_{1}})(\frac{z}{2 \delta K_{1}}) 
   \sinh{(\delta K_{1} z)}+
\frac{\delta^2 k_{x}^2}{2 K_{1}^2}\frac{z^3}{3} \sinh{(\delta K_{1}z)}- \\
 \frac{\delta^2 k_{x}^2}{2 K_{1}^2}(z^2/2 \delta K_{1})\cosh{(\delta K_{1} z)}+ 
\frac{\delta^2 k_{x}^2}{2 K_{1}^2}(z/2 \delta^2 K_{1}^2)
  \sinh{(\delta K_{1} z)}\}
+ G_{0 \zeta Y}(\frac{i \delta k_{x}}{K_{1}})(\frac{i \delta k_{y}}{K_{1}})[\mu^2 z^2 \cosh{(\delta K_{1}z)} -\\
\mu^2
\sinh{(\delta K_{1} z)} (\frac{z}{ \delta K_{1}})
]+ G_{11Y}[-(\frac{i \delta k_{y}}{K_{1}})\mu^2 z \sinh{(\delta K_{1} z)}]
$

\end{document}